\documentclass[10pt]{iopart}

\usepackage{iopams}  \usepackage{amssymb}
\usepackage{rawfonts}
\usepackage[usenames,pctex32]{color}

\input prepictex
\input pictex
\input postpictex

\def\a#1{{\bf a}_#1}

\def\d#1{{\bf d}_#1}
\def\dd#1#2{{\bf d}_{#1#2}}

\def\b#1{{\bf #1}}
\def\c#1{{\bf c}_{#1}}
\def\u#1{{\bf u}_{#1}}
\def\e#1{{\bf e}_{#1}}
\def\v#1{{\bf v}_{#1}}

\newcommand{\Zm}{{\mathbb Z}}
\newcommand{\Nm}{{\mathbb N}}
\newcommand{\Rm}{{\mathbb R}}

\newcommand{\Lm}{{\mathbb L}}

\newcommand{\U}{\underline}

\newtheorem{theorem}{Theorem}%[section]
\newtheorem{lemma}[theorem]{Lemma}

\newcommand{\Ref}[1]{(\ref{#1})}
\newcommand{\Proof}{\noindent {\it Proof: }}
\newcommand{\qed}{\hfill $\diamondsuit$}
\usepackage{multirow}
\usepackage{graphicx}
\newcommand{\ds}{\displaystyle}

\begin{document}

\title[Ergodic Proofs]{BFACF-style algorithms for polygons in the
body-centered and face-centered cubic lattices}

\author{E J Janse van Rensburg$\dagger$\footnote[3]{To whom 
correspondence should be addressed (\texttt{rensburg@yorku.ca)}}
and A Rechnitzer$\ddagger$}

\address{$\dagger$Department of Mathematics and Statistics, 
York University\\ Toronto, Ontario M3J~1P3, Canada\\
\texttt{rensburg@yorku.ca}}

\address{$\ddagger$Department of Mathematics, 
The University of British Columbia\\
Vancouver V6T~1Z2, British Columbia , Canada\\
\texttt{andrewr@math.ubc.ca}}

\begin{abstract}
In this paper the elementary moves of the BFACF-algorithm 
\cite{AC83,ACF83,BF81} for lattice polygons are generalised 
to elementary moves of BFACF-style algorithms for
lattice polygons in the body-centred (BCC) and face-centred
(FCC) cubic lattices. We prove that the ergodicity classes 
of these new elementary moves coincide with the knot
types of unrooted polygons in the BCC and FCC lattices 
and so expand a similar result for the cubic lattice 
(see reference \cite{JvRW91}). Implementations of these 
algorithms for knotted polygons using the GAS algorithm 
produce estimates of the minimal length of knotted polygons 
in the BCC and FCC lattices.
\end{abstract}

\maketitle

\section{Introduction}

\begin{figure}[t]
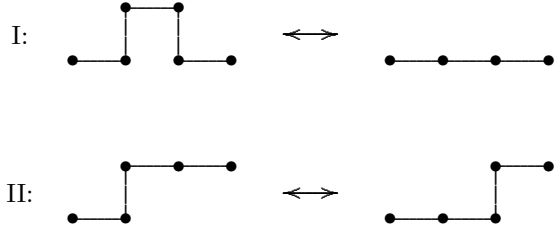

\input figa4.TEX
\caption{BFACF moves on a part of a square or simple 
cubic lattice polygon. Moves of type~I are positive 
elementary moves when 2 edges are added, and negative 
elementary moves when 2 edges are removed. Moves of type~II 
do not change the length of the polygon, and are called 
neutral elementary moves.}
\label{figa4} %%ZXZ[figa4]
\end{figure}

The BFACF algorithm \cite{AC83,ACF83,BF81} is a Metropolis 
style Monte Carlo algorithm \cite{MRRTT53} that samples 
self-avoiding walks with fixed endpoints in the grand 
canonical ensemble in the hypercubic lattice~$\Zm^d$. 
The algoritm uses the local elementary moves in 
Figure~\ref{figa4} to sample walks along a Markov Chain 
from the (non-Boltzman) distribution
\begin{equation}
P_\beta = \frac{\ds n\,e^{\beta n}}
{\ds \sum_{n\geq 0} n\, c_n (\mathbf{x},\mathbf{y}) 
e^{\beta n}}
\label{GC} %%ZXZ[GC]
\end{equation}
where $\beta$ is a parameter and $c_n(\mathbf{x},\mathbf{y})$
is the total number of self-avoiding walks of length $n$ 
from the lattice site $\mathbf{x}$ to the lattice site 
$\mathbf{y}$.

%%%%%%%%%%%%%%%%%
%%%%%%%%%%%%%%%%

The BFACF algorithm has also been used to sample unrooted 
polygons in the square and cubic lattices (see for example 
references \cite{BOS07,BOW09,JvROSTW96,JvRP95,OJ98,Q94} and 
it has been generalised to lattice ribbons \cite{OJvRW96A}. 

The BFACF algorithm can be used to sample knotted lattice
polygons of fixed knot type \cite{JvRW91} (also see 
\cite{JvR09} for a review) and has recently been used
to determine the entropy and length of minimal knots in 
the cubic lattice \cite{DIASV09,UJvROTW96}. 

Despite its widespread use, the BFACF algorithm appears to 
have only been used in  simple hypercubic lattices . In 
this paper our aim is to extend the elementary moves
of the BFACF algorithm (see Figure~\ref{figa4}) 
in the simple cubic lattice (hereafter referred to as the
SC) to the body-centred and face-centred cubic lattices 
(hereafter referred to as the BCC and FCC lattices 
respectively).  In addition, we examine the ergodicity 
properties of the proposed elementary moves when applied to 
unrooted self-avoiding polygons in the BCC and FCC lattices.

The BFACF algorithm is known to have elementary moves with
non-trivial ergodicity properties in the cubic lattice 
\cite{JvRW91,JvR92}.  In this paper we prove an analogous
result for the BCC and FCC lattices.

BFACF moves on walks or polygons (see Figure~\ref{figa4})
have been generalised by re-interpretation as 
{\it plaquette atmospheres} (see \cite{JvRR08}) --- 
namely the ways in which edges can be added, deleted 
or shuffled around plaquette\footnote[1]{Unit 
squares in the lattice bounded by four lattice edges} 
adjacent to edges in the polygon.  Examining the set of
such possible moves has proved extremely useful (see 
\cite{JvRR08}) and inspired generalisations 
of the Rosenbluth algorithm \cite{RR55} to the GARM 
\cite{RJvR08} and GAS algorithms \cite{JvRR09, JvRR10}. 

There are natural analogues of these plaquette atmospheres 
on the BCC and FCC lattices and the new elementary moves 
follow immediately from the definitions
of these atmospheres; see Sections~\ref{section31} 
and~\ref{section41}. While these moves are quite easy 
to define, some work is required to demonstrate their
ergodicity properties.

In reference \cite{JvRR10} an implementation of the 
BFACF elementary moves using the GAS algorithm was 
used to determine the shortest knots of given type in the 
simple cubic lattice. In this paper we extend those 
results by adding data for the BCC and FCC lattices, 
implementing the new BFACF-style elementary moves using 
the GAS algorithm.  The minimal lengths of a selection
of knot types in these lattices are displayed in
Table~\ref{table}.  Knot types are indicated in the 
first column, while the next three columns give the 
lengths of the shortest knots of given type in each 
lattice.  The last three columns are the number of 
distinct polygons (or ``population") of the lattice knots
of minimal length.

%%%%%%%%%%%%%%%%%
\begin{table}[h]
\caption{Minimal Knots in Cubic Lattices}
\label{table}
\begin{center}
\begin{tabular}{| c | r r r | r r r |}
% \hline & Minimal & Length & & Population & & \\ \hline
\hline & \multicolumn{3}{|c|}{Minimal  Length} & \multicolumn{3}{|c|}{Population} \\ \hline
Knot & SC & BCC & FCC & SC & BCC & FCC \\ \hline
$0_1$ & 4 & 4 & 3 & 3 & 12 & 8 \\ \hline
$3_1$ & 24 & 18 & 15 & 3328 & 1584 & 64 \\ \hline
$4_1$ & 30 & 20 & 20 & 3648 & 12 & 2796 \\ \hline
$5_1$ & 34 & 26 & 22 & 6672 & 14832 & 96 \\
$5_2$ & 36 & 26 & 23 & 114912 & 4872 & 768 \\ \hline
$6_1$ & 40 & 28 & 27 & 6144 & 72  & 19008  \\
$6_2$ & 40 & 28 & 27 & 32832 & 8256  & 5040 \\ 
$6_3$ & 40 & 30 & 28 & 35522 &  3312 & 102720  \\ \hline
$3_1^+\# 3_1^+$ & 40 & 30  & 26  & 30576 & 14520  & 960  \\
$3_1^+\# 3_1^-$ & 40 & 30  & 26  & 143904 & 24048  & 960  \\
\hline
\end{tabular}
\end{center}
\end{table}
%%%%%%%%%%%%%%

In Section~\ref{section2} we review the ergodicity 
properties of the BFACF elementary moves in the simple 
cubic lattice.  We recall that the irreducibility classes of
these moves, when applied to unrooted simple cubic lattice 
polygons, are the knot types of the polygons as embeddings 
of the circle in $\Rm^3$.  This result, in particular,
implies that two unrooted cubic lattice polygons $\omega_1$ 
and $\omega_2$ are in the same knot type if and only if 
there is a sequence of (reversible) BFACF elementary moves \
which will take $\omega_1$ to $\omega_2$ \cite{JvRW91}. 
We generalise this result to the BCC and the FCC lattices.

In Section~\ref{section3} we define an analogous set of 
elementary moves for polygons in the BCC lattice. We examine 
the properties of these moves and prove that the projection of
any (unrooted) BCC polygon can subdivided by application of the 
BCC elementary moves.  A corollary of this result is that 
polygons can be made contact free --- roughly speaking
the polygon can be ``inflated'' so that non-consecutive vertices 
do not lie close to each other. This result is then used to 
sweep the BCC polygon into a sublattice isomorphic to
the simple cubic lattice. In this sublattice the BCC 
elementary moves reduce to the usual BFACF moves. This 
is sufficient to show that the BCC elementary moves are
irreducible on the classes of unrooted BCC polygons of 
given knot type.

Polygons in the FCC lattice are examined in 
Section~\ref{section4}. In this lattice the analogous of 
the cubic lattice BFACF moves is a single reversible elementary 
move which increases or decreases the length of an FCC 
lattice polygon by one. The proof proceeds similarly to the 
BCC case, albeit with more cases. Again we show that any FCC 
lattice polygon can be made contact free and be swept onto 
a sublattice isomorphic to the simple cubic lattice.  The 
usual set of simple cubic lattice BFACF moves can be performed 
on the polygon (in this sublattice) by compositions of the 
FCC elementary move. Similarly, this is sufficient to show 
that the FCC elementary moves are irreducible on the classes 
of unrooted FCC polygons of given knot type.

In Section~\ref{section5} we show how the elementary moves 
may be used in a BFACF-style algorithm.  This 
implementation samples from the distribution of polygons 
similar to that in equation~\Ref{GC}. In addition, we note 
that the elementary moves can be implemented using the 
Metropolis algorithm instead, or using GAS-style sampling. 
We conclude the paper with a few final comments.

%%%%%%%%%%%%%%%%%%%%%%%%%%%%%%%%%%%%%%%%%%%%%%%%%%%%%%%%%%%%%%%
%%%%%%%%%%%%%%%%%%%%%%%%%%%%%%%%%%%%%%%%%%%%%%%%%%%%%%%%%%%%%%%
\section{Knotted Lattice Polygons}
\label{section2}

Let $S$ be the circle; we consider an injective map 
$f:S\to\Rm^3$, to be an {\it embedding} of the circle in 
Euclidean three space (that is, $f$ is an injection, and is a
homeomorphism onto its image).  A polygon in the simple 
cubic lattice (SC), or the BCC lattice, or the FCC lattice, 
is a piecewise linear embedding of $S$ into $\Rm^3$. Any 
such embedding of $S$ is a {\it knot}, and if the embedding is 
a lattice polygon, then the embedding is a {\it lattice knot}. 
In this way cubic lattice polygons are lattice knots.

Two oriented embeddings $f$ and $g$ are ambient isotopic if 
there is an orientation-preserving isotopy $H: \Rm^3\times I 
\to \Rm^3\times I$ (where $I=[0,1]$) with $H(y,t) 
\equiv (h_t(y),t)$ such that $h_0$ is the identity 
($h_0 \circ f=f$) and the composition $h_1\circ f = g$.  
In other words, two lattice polygons are ambient isotopic
if there is a continuous deformation of $\Rm^3$ which takes 
the embedding $f$ of the first polygon onto the embedding $g$ 
of the second polygon.

Two lattice polygons in any cubic lattice (ie the SC, BCC or 
FCC lattices) are said to be equivalent if they are ambient 
isotopic.  These equivalence classes of oriented embeddings
of the circle into the cubic lattices define the knot types 
of {\it lattice knots}, see for example references 
\cite{JvR08} for a reviews and definitions of lattice knots. 

In this paper we prove that there exists piecewise linear 
realisations of orientation-preserving ambient isotopies 
between lattice knots of the same knot type in the BCC and 
FCC lattices. These isotopies can be constructed as sequences 
of local deformations of the lattice knots in terms of 
BFACF-style elementary moves. This is an extension of a 
similar result for polygons on the simple cubic lattice in 
\cite{JvRW91}. In that result, the piecewise linear 
orientation-preserving isotopies are realised in steps using 
the elementary moves of the BFACF algorithm, illustrated 
in Figure~\ref{figa4}. By constructing the isotopies in this 
way in reference \cite{JvRW91} the following result is proven:

%%%%%%%%%%%%%%%%%%%%%%%%%%%%%%%%%%%%%%%%%%%%%%%%%%%%%%%%%%%
\begin{theorem}[Thm 3.11 from \cite{JvRW91}]
The irreducibility classes of the BFACF algorithm, when 
applied to unrooted polygons [in the simple cubic lattice], 
are the knot types of the polygons as piecewise linear 
embeddings in $\Rm^3$. 
\label{thm 11} %%ZXZ[thm 11]
\qed
\end{theorem}
%%%%%%%%%%%%%%%%%%%%%%%%%%%%%%%%%%%%%%%%%%%%%%%%%%%%%%%%%%%

This, in particular, follows by proving that for every pair 
of (oriented) polygons $(\omega,\nu)$ of the same knot type, 
there exists a finite sequence of elementary moves of either 
type~I or~II in Figure~\ref{figa4}, such that these moves 
change $\omega$ into $\nu$.  Each move is a local deformation 
of the polygon and the ambient space around it, and is itself 
an isotopy.  The sequence of moves is a composition of these 
local isotopies, and is itself a realisation of an isotopy 
$H: \Rm^3\times I \to \Rm^3\times I$ such that $H(y,t) 
\equiv (h_t(y),t)$ where $h_0$ is the identity and 
$h_1\omega = \nu$.

In other words, the equivalence classes of polygons induced 
by the simple cubic lattice BFACF elementary moves in 
Figure~\ref{figa4} coincides with the knot types of the
polygons, proving that the irreducibility classes of the 
BFACF algorithm are the knot types of unrooted polygons 
in the simple cubic lattice. In this paper, we extend this
result to the BCC and FCC lattices --- these extensions 
are Theorems~\ref{thm BCC main} and~\ref{thm FCC main} below.

% In particular, we shall 
% prove that there exist piecewise linear realisations 
% of orientation-preserving ambient isotopies 
% between lattice knots of the same knot type in the BCC
% and FCC lattices, and we give explicit realisations of these
% isotopies in terms of local deformations of the lattice
% knots in terms of BFACF-style elementary moves of 
% Monte Carlo algorithms for polygons in the BCC and 
% FCC lattices.

\subsection{Lattice Knot Projections}

The BCC and FCC lattices are generated by finite sets of 
basis vectors $\{\c{i} \}$ and $\{\e{i} \}$.  Once the
origin $O$ of the lattice is set, then {\it vertices}
are defined by linear combinations $\sum_i p_i \c{i}$
and $\sum_i p_i \e{i}$ respectively in the BCC and FCC, where
the $p_i \in \Zm$ are finite signed integers.  Thus,
in each case the vertices have integer Cartesian coordinates.

Two vertices $\u{}$ and $\v{}$ are adjacent if the difference
$\u{}-\v{}$ is a basis vector of the lattice.  In this
case an {\it edge} $\u{}\v{}$ is defined between the
vertices. The vertices $\u{}$ and $\v{}$ are the 
{\it end-vertices} of the edge $\u{}\v{}$. We consider 
a lattice to be the collection of
all its vertices and edges.  Two vertices are adjacent if
they are the endpoints of the same edge.  Two edges are
adjacent (or incident on one another) if they share exactly
one end-vertex.

A {\it lattice polygon} is defined as a sequence or list
of $n$ adjacent edges $\langle \u{0}\u{1},\u{1}\u{2},\ldots, 
{\bf u}_{n-1}\u{0} \rangle$, such that all vertices 
$\{\u{0},\u{1},\ldots,{\bf u}_{n-1}\}$ are distinct. 

The {\it length} of the polygon is the number of edges 
$n$ it contains (but its geometric length will generally
be different from this, since the edges do not necessarily
have length equal to one).

The BCC lattice is Eulerian with girth $4$, and all 
lattice polygons in it have even length. The FCC lattice
is Eulerian of girth $3$, and polygons of length longer
or equal to $3$ can be realised in this lattice.

A lattice edge $\u{i}\u{j}$ in the BCC lattice is said 
to be parallel to the $\c{i}$ direction, or {\it in the 
$\c{i}$ direction} if $\u{j} - \u{i} = \c{i}$.  Similarly,
one may define edges to be in the $\e{i}$ direction in
the FCC lattice. When a lattice edge in the BCC is parallel 
to the $\c{i}$ direction (or to the $\e{i}$ direction in 
the FCC), then we shall frequently abuse our notation 
by denoting it by its direction $\c{i}$ (or by $\e{i}$).

A {\it line segment} in a lattice polygon is a maximal 
non-empty sequence of adjacent edges in the polygon of 
the form $\c{i}\c{i}\ldots\c{i}$ (in the BCC lattice),
or $\e{i}\e{i}\ldots\e{i}$ (in the FCC lattice). 
We say that these line segments are in the $\c{i}$ of $\e{i}$
directions respectively.

In what follows, we shall work with the {\it projections}
of lattice polygons $\omega$into geometric planes $A$ 
along a direction $\u{}$. To define these projections, 
consider two independent (unit) vectors co-planar with $A$, 
and let $\b{z} = \b{x}\times\b{y}$ be a vector normal to 
$A$. A vector $\b{u}$ is {\it transverse} to $A$ if 
$\b{u}\cdot \b{z} \neq 0$. 

The three vectors $\{ \b{x},\b{y},\b{u} \}$ is the basis of
a (non-orthogonal) coordinate system $S$ in $\Rm^3$.
Points in the polygon $\omega$ can be identified by 
their coordinates in $S$, for example, $\omega$ is a 
piecewise linear curve parametrised by $t$ and each 
point $\omega(t)$ has coordinates $(\b{x}_t,\b{y}_t,\b{u}_t)$.

The {\it projection} of $\omega$ into $A$ along $\bf{u}$ 
is defined by the set of points $(\b{x}_t,\b{y}_t)$ in $A$ 
for all values of the parameter $t$.

A {\it multiple point} in the projection of $\omega$ into 
$A$ along $\bf{u}$ is a point in the projection which is 
the image two or more distinct points in $\omega$.  A 
multiple point is a {\it double point} if it is the image 
of exactly two points.

In the case of lattice polygons, projections of polygons will be subgraphs of the projection of the lattice into a plane
normal to a given lattice axes. For example, the projection 
of a simple cubic lattice polygon along the $Z$-direction 
into the $XY$-plane is a square lattice, and a cubic 
lattice polygon will project into a subgraph of this square 
lattice.  This projection is a {\it lattice knot projection} 
in the square lattice; see reference \cite{JvRW91}. 

In the BCC and FCC lattices we shall take projections 
of lattice polygons onto symmetry planes of the lattice, 
along directions which are transverse but not necessarily
orthogonal to the symmetry planes. 

%%%%%%%%%%%%%%%%%%%%%%%%%%%%%%%%%%%%%%%%%%%%%%%%%%%%%%%%%%
%%%%%%%%%%%%%%%%%%%%%%%%%%%%%%%%%%%%%%%%%%%%%%%%%%%%%%%%%%
%%%%%%%%%%%%%%%%%%%%%%%%%%%%%%%%%%%%%%%%%%%%%%%%%%%%%%%%%%
\section{BFACF-Style Elementary Moves in the BCC Lattice}
\label{section3}

In this section, we propose the local elementary moves 
of a BFACF-style algorithm in the BCC lattice and we show 
that they are sufficient to realise a
piecewise linear orientation preserving isotopy on 
unrooted BCC polygons of the same knot type embedded in 
$\Rm^3$.  In particular, this implies that the 
irreducibility classes of the BCC elementary moves coincide 
with the knot types of unrooted BCC polygons as determined 
by their embeddings in three space.

\begin{figure}
 \begin{center}
  \includegraphics[height=6cm]{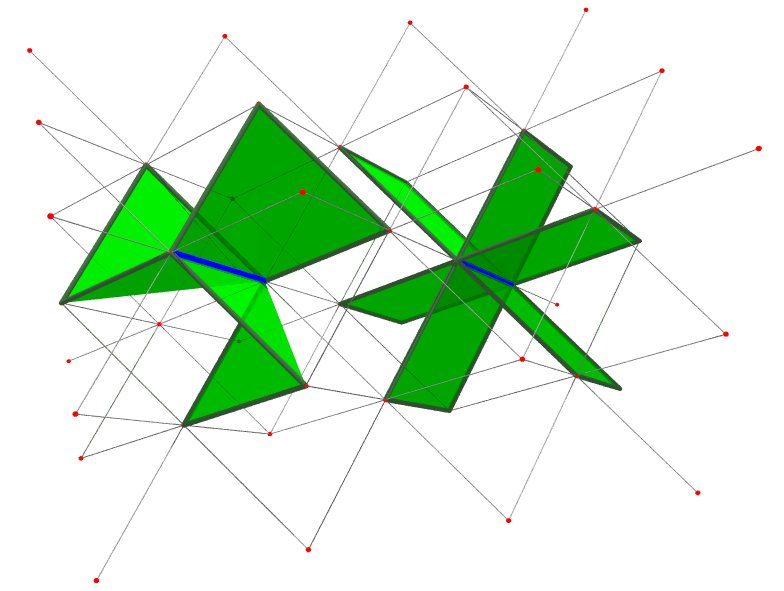}
 \end{center}
  \caption{
Nine of the 12 plaquettes adjacent to an edge in the BCC
lattice. Note that the six on the right are planar, while 
the three on the left are non-planar. The remaining 
three (undisplayed) polygons are mirror-images of the 
three non-planar polygons.}
  \label{fig bcc plaquettes}
\end{figure}

\subsection{BFACF style moves in the BCC lattice}
\label{section31}

In section 2.1 the notion of lattice polygons and projections
of polygons were defined in general. We note in particular
that the basis vectors of the BCC lattice are points in 
$\Rm^3$ with Cartesian coordinates given by 
$p\c1+q\c2+r\c3+s\c4$ where $p,q,r,s\in\Zm$, and where 
the vectors $\c{i}$ are given by
\begin{displaymath}
\hspace{-8ex}
\begin{array}{llll}
\c1 = (1,1,1),& \c2 = (1,1,-1), & \c3 = (1,-1,1), & \c4 = (1,-1,-1), \\
\c5  = (-1,-1,-1),& \c6 = (-1,-1,1),& \c7 = (-1,1,-1),& \c8 = (-1,1,1) .
\end{array}
\end{displaymath}
Observe that $\c5=-\c1$, $\c6=-\c2$, $\c7=-\c3$ and 
$\c8=-\c4$. The vectors $\c{i}$ the {\it generating} or 
{\it basis vectors} of the BCC and they all have
(geometric) length $\sqrt{3}$.

Two vertices, $\b{u}$ and $\b{v}$, are {\it adjacent}
if they are separated by a single basis vector $\c{i}$.
Lattice edges will be in the $\c{i}$ directions in the
BCC, and each lattice edge $\c{i}$ will be an edge in 
minimal length polygons of length $4$ in the BCC lattice.
These minimal length polygons containing $\c{i}$ bound
$12$ faces in the BCC, and these faces are not square and 
may not be planar in $6$ of the cases.  Nevertheless, we 
shall refer to them as lattice {\it plaquettes} -- see 
Figure~\ref{fig bcc plaquettes}.

\begin{figure}[t]
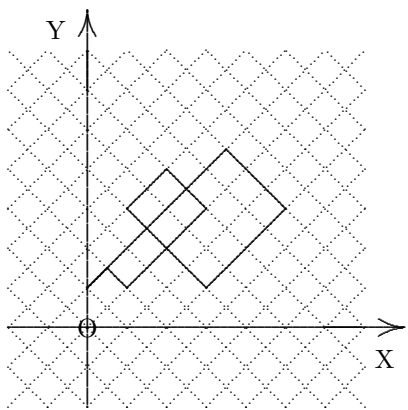

\input figa1.TEX
\caption{A projection of the BCC lattice and a BCC lattice
polygon into the $XY$-plane.  The projection of the BCC lattice
is a square lattice rotated at $45^o$ degrees with the 
Cartesian axes, and with edge-length $\sqrt{2}$. The polygon
projects to a subgraph of the projected lattice.}
\label{figa1} %%ZXZ[figa1]
\end{figure}

The (orthogonal) projection of the BCC lattice into the\
$XY$-plane is a geometric square lattice with edge lengths 
$\sqrt{2}$ and rotated at $45^o$ with respect to the $X$-axis.  
A polygon in the BCC projects to a subgraph of this 
square lattice.  This is illustrated in Figure~\ref{figa1}.

%%%%%%%%%%%%%%%%%
%%%%%%%%%%%%%%%%

We define elementary moves on polygons in the BCC lattice 
by considering the $12$ plaquettes adjacent to every edge 
$\c{i}$ (see Figure~\ref{fig bcc plaquettes}). Collectively,
these plaquettes composed the {\it plaquette atmosphere} of
the polygon \cite{JvRR08}.  In the 
simple cubic lattice, each edge is incident on at most
$4$ atmospheric plaquettes, and elementary moves of the BFACF
algorithm are obtained by selecting an atmospheric plaquette
and then exchanging edges along it boundary to update the 
polygon.  

In particular, by taking the alternate path around
the boundary of atmospheric plaquette BFACF moves in 
Figure~\ref{figa4} is obtained. A type~I move on the 
simple cubic lattice performed by replacing a single
edge incident on an atmospheric plaquette $P$ in the polygon 
by the $3$ edges in the alternative path around the boundary
of $P$. This move is reversible, and together the move and
its reverse defines type~I moves. Similarly, a type~II
move on the cubic lattice replaces a pair of edges on 
incident on an atmospheric plaquette $P$ with the other two 
edges in $P$.

\begin{figure}[t]
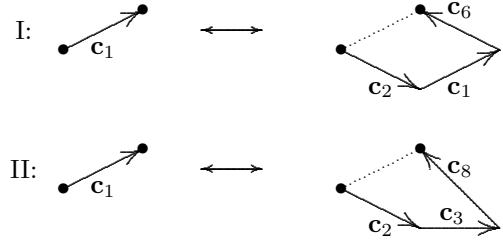

\input figa2.TEX
\caption{Replacing $\c1$ by three edges as shown defines 
an elementary move which increases the length of a polygon in 
the BCC lattice by two.  This defines a {\it positive
plaquette atmospheric move} on the polygon.  Reversing the 
move gives a {\it negative plaquette atmospheric move} 
which reduces the length of the polygon by two edges by 
replacing three edges on the right by the single edge on 
the left.  All the possible permutations of the vectors
$\c{i}$ gives the complete collection of positive and 
negative plaquette atmospheric moves.  Observe that closing 
the vectors on the right with the dotted line gives a square 
in case I, and a non-planar quadrilateral in case II.}
\label{figa2} %%ZXZ[figa2]
\end{figure}

The $12$ atmospheric plaquettes incident on edges in BCC 
polygons will similarly be used to determine the set of 
elementary moves on BCC lattice polygons.  A similar analysis
to the above gives the following set of elementary moves 
on polygons in the BCC lattice:
 \vspace{5mm}\\
\noindent{\U{BCC Elementary Moves:}}
\begin{itemize}
\item Replace an edge $\c1$ by $\c2\c1\c6$ (and all 
permutations of these vectors in the set of vectors $\c{i}$).
This move increases the length of the polygon by $2$. 
Conversely, replace $\c2\c1\c6$ by $\c1$ (and all permutations 
of these vectors) to obtain a move reducing the length of 
the polygon by $2$. This moves is illustrated (generically)
by case~I in Figure~\ref{figa2}. Observe that the vectors 
$\{\c2,\c1,\c6\}$ are coplanar, so that $\c1\c2\c1\c6$ forms 
a planar quadrilateral --- a square. Hence, we call this the 
{\it planar BCC positive and negative plaquette atmospheric 
moves}, or the {\it planar BCC positive and negative 
elementary moves}.
\item Replace an edge $\c1$ by $\c2\c3\c8$ (and all 
permutations of these vectors).  This move increases the 
length of the polygon by $2$.   Conversely, replace 
$\c2\c3\c8$ by $\c1$ (and all permutations of these vectors) 
to obtain a move reducing the length of the polygon by $2$. 
This move is illustrated by case II in Figure~\ref{figa2}. 
Observe that the set of vectors $\{\c2,\c3,\c8\}$ is 
not coplanar; hence closing them off with $\c1$ bounds 
a non-planar quadrilateral or atmospheric plaquette. These 
are the {\it non-planar BCC positive and negative 
plaquette atmospheric moves}, or the {\it non-planar BCC 
positive and negative elementary moves}.
\item Replace $\c1\c6$ by $\c6\c1$ (and all other 
permutations of these vectors).  This is a neutral move, 
illustrated as case I in Figure~\ref{figa3}. Note that these 
two pairs of edges form a square and so this move occurs 
in a plane in three space. These are the {\it planar BCC 
neutral plaquette atmospheric moves}, or the {\it planar 
BCC neutral elementary moves}.
\item Replace $\c1\c6$ by $\c3\c2$ (all other 
permutations of these vectors). This is a neutral 
move, illustrated as case II in Figure~\ref{figa3}. Now, 
these two pairs of edges form a non-planar quadrilateral 
or atmospheric plaquette and so this move does not occur 
in a plane in three space. More generally, these are the 
{\it non-planar BCC neutral plaquette atmospheric moves}, 
or the {\it non-planar BCC neutral elementary moves}.
\end{itemize}

%%%%%%%%%%%%%%%%%
%%%%%%%%%%%%%%%%

Projecting the moves of Figures~\ref{figa2} and~\ref{figa3} 
into the $XY$-plane gives the (two dimensional) square 
lattice BFACF moves shown in Figure~\ref{figa4}, but in the
rotated square lattice of Figure~\ref{figa1} instead.  Thus, 
by executing the BCC positive, negative and neutral 
plaquette atmospheric moves of Figures~\ref{figa2} 
and~\ref{figa3} on a polygon in the BCC, the image of these 
moves in the projected square lattice include, as a subset, 
all the (usual simple square lattice) BFACF moves on the 
projected polygon.

%%%%%%%%%%%%%%%%%
\begin{figure}[t]
\input figa3.TEX
\caption{Replacing $\c6\c1$ by either I: $\c1\c6$ or by II: 
$\c3\c8$ defines neutral plaquette atmospheric moves.  
Observe that $\c1+\c6 = \c3+\c8$.  All the other possible 
neutral moves are obtained by using all the possible 
permutations of the $\c{i}$ in the above.  Observe that 
by closing the vectors on the right into a quadrilateral 
along the dotted lines (which follows the path of $\c6\c1$ 
between the bullets) gives a square in case I and a 
non-planar quadrilateral in case II.}
\label{figa3} %%ZXZ[figa3]
\end{figure}
%%%%%%%%%%%%%%%%

%%%%%%%%%%%%%%%%%%%%%%%%%%%%%%%%%%%%%%%%%%%%%%%%%%%%%%%%%%%%%%%%
\subsection{Stretching polygons in the BCC lattice}

We first outline our approach before we prove our main results.
The proof of our main BCC lattice result in 
Theorem~\ref{thm BCC main}, is presented in two parts. 
We first show that any BCC lattice polygon can be swept onto 
a sublattice of the BCC which is isotopic to the simple 
cubic lattice (as an oriented embedded graph in three space). 
Then we show that in this sublattice we use the 
ergodicity properties of the simple cubic lattice BFACF
algorithm \cite{JvRW91} to complete the proof 
(see Theorem~\ref{thm 11}).

In other words, we shall show that any BCC lattice polygon
can be swept into a sublattice $\Lm$ with basis vectors 
$\{\c1,\c3,\c4,\c5,\c7,\c8\}$, and then show that a subset of the
BCC elementary moves simulates a simple cubic lattice BFACF
algorithm in this sublattice (which is not orthogonal but 
is nevertheless isotopic to the simple cubic lattice).

Our approach would be to demonstrate that the BCC elementary
moves are sufficient to replace polygon edges outside 
the sublattice $\Lm$ with edges in $\Lm$, while avoiding any
self-intersections in the polygon as it is updated in this process.
This is achieved by {\it stretching} the polygon to create
sufficient space for executing BCC elementary moves
(see Figure~\ref{figa6}).

The stretching of a polygon will proceed by identifying
a {\it maximal line} which intersect the projected polygon in 
its right-most and top-most projected edges.  The polygon will
be recursively stretched in directions normal to the maximal 
line by stretching parts of it across the maximal, while 
inserting edges to maintain its connectivity.  We show
that this can be done using the BCC elementary moves.

In particular, project the BCC lattice along the 
$Z$-direction onto the $XY$-plane, and let $\omega$ be a 
BCC polygon with projection $P\omega$ in the $XY$-plane.  
Then the image of the lattice and the polygon is the 
square lattice and a graph embedded in the square lattice
illustrated in Figure~\ref{figa1}. The {\it maximal} line $K$
of $P\omega$ is the line $x+y=k$ with $k$ the maximum value
such that $K$ has a non-empty intersection with $P\omega$.

$K$ is the image of a plane $P_r$ projecting parallel to the 
$Z$-axis into the $XY$-plane.  We say that $P_r$ is maximal 
if it intersects $\omega$ and if it projects to the maximal
line.

The maximal line $K$ intersects the projection $P\omega$ 
in projected line segments which lifts to line segments 
and isolated vertices in $\omega \cap P_r$.

These definitions can now be used to stretch a polygon in
a consistent way without creating self-intersections.

\subsubsection{Stretching a polygon in its maximal line:} 
Suppose $L$ is a line given by $x+y=l$ (with $l\in \Nm$) 
which intersects the projected polygon $P\omega$. The 
maximal line $K$ has equation $x+y = k$, and necessarily, 
$k \geq l$. 

\begin{figure}[t]
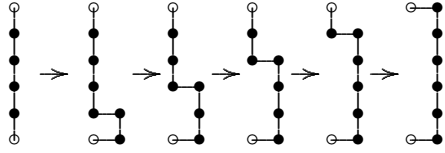

\input figa6.TEX
\caption{Translating a line segment with a single positive 
move and then a sequence of neutral moves in the BCC lattice, 
projected into the $XY$-plane.  Since the projection
of the polygon is to a square lattice, the BCC elementary 
moves project to square lattice BFACF elementary moves.}
\label{figa6} %%ZXZ[figa6]
\end{figure}

We say BCC basis vectors $\c{i}$ are {\it transverse} to 
the plane $P_r$ if they are not parallel to $P_r$. Let $\c{i}$ 
be transverse to $P_r$ ---  
then $\c{i} \in \{\c{1},\c2,\c5,\c6\}$; 
the other lattice vectors lie in $P_r$. Each line segment 
in this intersection can be translated one step in the 
$\c{i}$ direction by BCC elementary moves using the 
construction in Figure~\ref{figa6}. We restrict such 
translations to be in the $\c{i}$ direction (with $i=1$ or 
$i=2$) in what follows. This will translate all line segments of
non-zero length in the plane $P_r$ in the $\c{i}$ direction.

This leaves the case of isolated vertices in $P_r\cap \omega$.
The edges incident to and from these vertices will be of the form
$\c{1}\c{6}$ or $\c{2}\c{5}$ (since $K$ is maximal). In this 
case one may perform the move 
$\c{1}\c{6} \to \c{1}\c{1}\c{6}\c{5}$ or $\c{2}\c{5} \to
\c{2}\c{1}\c{5}\c{6}$ to translate the isolated vertex in the 
$\c{1}$ direction.  A similar construction will translate 
the segment in the $\c{2}$ direction instead.

Observe that no self-intersections can occur since all parts
of the polygon in $P_r\cap \omega$ are translated in parallel
in the $\c{i}$ direction (with $i=1$ or $i=2$).

Once all line segments in the maximal plane $P_r$ are translated in the $\c{1}$ (or $\c{2}$) direction, the polygon $\omega$ 
is said to have been {\it stretched} in the plane $P_r$ in 
the direction~$\c{1}$.

On completion of this construction, the maximal line $K$ is translated to the line $L$ by translating the plane $P_r$
in the $-\c{i}$ direction.  This plane is denoted by $Q_r$
and is parallel to $P_2$.  Then $Q_r$ projects to $L$ with
formally $x+y=\ell = k-1$ in the $XY$-plane. $L$ lifts 
to $Q_r$, and $Q_r \cap \omega$ is a collection of line segments
and isolated vertices of $\omega$.  

Observe that the departing edges from $Q_r$ to the maximal 
side arriving in $P_r$ are all in the $\c{i}$ direction, 
since the polygon was stretched in that direction in the 
previous step. 

In addition, the $P_r$ contains no line segments, since 
these were translated into the $\c{i}$ direction.  
Furthermore, edges incident arriving in $P_r$
from the opposite of the maximal side are parallel or 
anti-parallel to $\c{j}$ with $j=1$ or $j=2$.

%%%%%%%%%%%%%%%%%
%%%%%%%%%%%%%%%%

\subsubsection{Stretching a BCC polygon}

We proceed by recursively executing the stretching of $\omega$
in the $\c{i}$ direction in the plane $Q_r$.  The intersection
$Q_r \cap \omega$ is a collection of line segments and
isolated vertices in $\omega$.  

If $Q_r = P_r$, where $P_r$ projects to the maximal line $K$,
then the situation is as described in the last section:
We must consider two different cases in translating parts 
of the polygon in $P_r$ in the $\c{i}$ direction. 
The first case involves line segments in $P_r\cap\omega$, 
the second case involves isolated vertices in 
$P_r\cap \omega$.  Such a isolated vertices must have 
incident edges $\c{1}\c{6}$ or $\c{2}\c{5}$ (so 
as not to collide with $P_r$). These two cases were already 
dealt with above.

In the event that $Q_r \not= P_r$, suppose that the stretching
was recursively done starting in $P_r$ and moving the plane
in the $-\c{i}$ direction so that the last stretching was done 
in the $\c{i}$ direction in the plane $Q_r^\prime = Q_r + 
\c{i}$. Without loss of generality, one may suppose that
the stretching is done in the $i=1$ direction.  Then all
edges between $Q_r$ and $Q_r^\prime$ are parallel or
anti-parallel to $\c{1}$.

There are three different cases to consider.  The first 
case involves line segments in $Q_r\cap\omega$, and the second 
case involves isolated vertices in $Q_r\cap \omega$ with
incident edges $\c{1}\c{6}$ or $\c{2}\c{5}$. These two cases 
are done by translating the vertices and edges in the $\c{1}$
direction similarly to those line segments and isolated
vertices in the plane $P_r$ above:  Since line segments projects
to lines in the square lattice and BCC moves to BFACF moves
in the square lattice, these line segments can be translated
by applying BCC moves in the $\c{1}$ direction.  Translating
isolated vertices in the second case is similarly done.

The third and final case involves isolated vertices $\b{v}$
of the polygon in $Q_r$ with edges on either side of $Q_r$. 
Suppose that these vertices are to be moved in the 
$\c{1}$ direction. Then the arrangements must be 
$\c{1}\b{v}\c{1}$, or $\c{2}\b{v}\c{1}$; the middle vertex 
$\b{v}$ in these cases lies in $Q_r$, and the second edge 
moves from $Q_r$ to a plane $Q_r^\prime = Q_r + \c{1}$. 

In the cases $\c{1}\b{v}\c{1}$ the edges are left unchanged, 
since the departing edges to $Q_r^\prime$ are already in the 
$\c{1}$ direction.  The case $\c{2}\b{v}\c{1}$ is updated to 
$\c{1}\b{w}\c{2}$, with $\b{w}$ in the plane $Q_r^\prime$.
Observe that $\b{w}$ is always not occupied in $Q_r^\prime$
before the move, because all arriving edges from $Q_r$
to $Q_r^\prime$ are in the $\c{1}$ direction before the
moves are done.

In each case, these constructions give a polygon with 
departing edges from $Q_r$ to $Q_r^\prime$ in the $\c{i}$
direction. Observe that the relative orientation of edges 
are again maintained, and that the moves are possible without 
any self-intersections in $\omega$.

The implementation of the stretching is recursive.  Start
in the plane $Q_r = P_r$ which projects to the maximal
line, and stretch the polygon a number of times in the
$\c{i}$ direction transverse to $Q_r$.  Then define
$Q_r^\prime = Q_r$ and $Q_r \to Q_r - \c{i}$ recursively.
This will stretch the polygon any desired length in each
plane $Q_r$ intersecting parallel to $P_r$ without creating 
an intersection in the polygon.

The effect of the construction is to cut $\omega$ along a plane 
$Q_r$ and to move the two parts of $\omega$ on either side
of the polygon any number of steps in the $\c{i}$ direction
transverse to $Q_r$ apart while inserting edges parallel 
or anti-parallel to $\c{i}$ to reconnect it into a single polygon.

Observe that the construction does not change the knot type of 
$\omega$, and that it is the realisation of an ambient
isotopy on the complementary space of the polygon.  We say 
that we have {\it stretched} the polygon $\omega$ in the 
$\c{i}$ direction along the plane $Q_r$. This completes the
construction.

Observe that a similar construction will enable one to 
stretch $\omega$ in directions transverse to lattice 
planes in the $BCC$ transverse to any of the basis vectors
$\c{i}$ of the BCC.  This follows by exchanging the
lattice basis vectors, such that a rotation of the polygon
preserving its chirality, is done in the above analysis.
These observations complete the proof of the following theorem:

%%%%%%%%%%%%%%%
\begin{theorem}
Let $\omega$ be an (unrooted) polygon in the BCC lattice 
with projection $P\omega$ in the $XY$- $YZ$- or $XZ$-plane. 
Let $L$ be a line in the $AB$-plane with formula $A\pm B =
\ell$ where $(A,B)$ is one of $(X,Y)$, $(Y,Z)$ or $(X,Z)$.

Suppose $L$ intersects the projection $P\omega$ and that 
$\ell$ is an integer.   Suppose also that $L$ lifts to the 
plane $Q_r$ parallel to the normal of the projection 
plane. Then $\omega$ can be stretched along the plane 
$Q_r$ in a direction $\c{i}$ transverse to $Q_r$ by 
performing the BCC elementary moves in Figures~\ref{figa2} 
and~\ref{figa3} on $\omega$. \qed
\end{theorem}
%%%%%%%%%%%%%%

\begin{figure}[t]
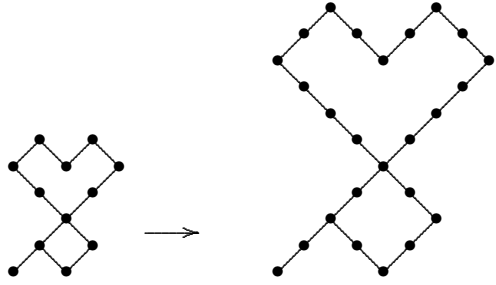

\input figa5.TEX
\caption{Stretching a projection of a BCC polygon. In this illustration, a polygon was
stretched recursively in the projected directions until each projected edge was doubled.}
\label{figa5} %%ZXZ[figa5]
\end{figure}

In Figure~\ref{figa5} we illustrate that a polygon can 
be recursively stretched until each projected edge has 
been replaced by two edges in the same projected direction.  
This is called the {\it subdivision} of the projected 
image of the polygon.

%%%%%%%%%%%%%%%%%
%%%%%%%%%%%%%%%%

\subsection{Contact free polygons}

In this section we show that a given BCC polygon can be
swept onto a sublattice $\mathbb{L}$ with basis vectors
$\{\c1, \c3, \c4, \c5, \c7, \c8\}$.  This sublattice is
isotopic to the simple cubic lattice, and once the polygon
is contained in it, then a subset of the BCC moves reduces
the usual BFACF moves.

In order to sweep a polygon $\omega$ into $\mathbb{L}$,
it is necessary to avoid self-intersections in the polygon
when executing the necessary elementary moves.  Such
possible self-intersections are avoided by stretching
the polygon, using the recursive subdivision explained in
the previous section.

Two vertices $\b{v}$ and $\b{w}$, non-adjacent in a polygon 
$\omega$ (ie not connected by an edge in $\omega$), form a 
{\it contact} if $\b{v}-\b{w} = \c{j}$ for some value of $j$. 

Such a contact is said to be a {\it contact in the direction 
$\c{j}$}.  A polygon $\omega$ is {\it contact free} if it has 
no contacts.  Observe that every contact is a lattice edge 
and projects to a line segment in the $XY$-plane.  We show
that by subdividing a polygon using the constructions outlined
in the previous sections, one can make it contact free.

%%%%%%%%%%%%%%%
\begin{lemma}
By applying the BCC elementary moves in Figures~\ref{figa2} 
and~\ref{figa3} to an unrooted polygon in the BCC lattice, 
it can be transformed into a contact free polygon.
\label{lemma3}
\end{lemma}
%%%%%%%%%%%%%%

\Proof
Use BCC elementary moves to remove all contacts in the 
polygon $\omega$ as follows.  

If $\omega$ has contacts in the $\c{i}$ direction, then 
subdivide the polygon such that the stretching is done 
in the $\c{i}$ direction.  Since all edges in the $\c{i}$ 
direction will be doubled up (effectively the scale of the
polygons is changed such that every edge is replaced by two
edges in a line segment of length $2$), this removes all 
contacts in the $\c{i}$ direction. If there are still
contacts in the $\c{j}$ direction (for some other $j$), 
then subdivide the polygon in the $\c{j}$ direction.

Observe that if there no contacts in the $\c{i}$ direction, 
then stretching the polygon in the $\c{j}$ direction 
cannot create new contacts in the $\c{i}$ direction.

This follows because the creation of a contact in the 
$\c{i}$ direction will require the translation of parts
 of the polygon relative to one another in the $\c{i}$
direction, which does not occur when stretching in the 
$\c{j}$ direction.

Repeat the process of subdivision, until all contacts are removed 
and a contact free polygon is obtained. \qed

\subsection{Pushing contact free BCC polygons into a 
simple cubic sublattice}

Our goal is to show that every polygon in the BCC lattice 
can be changed into a polygon in sublattice $\mathbb{L}$ which
is isotopic to the simple cubic lattice by the applying the 
BCC elementary moves. 

We defined $\mathbb{L}$ to be that sublattice of the BCC 
with basis vectors $\{\c1,\c3, \c4, \c5, \c7, \c8\}$.  
$\mathbb{L}$ is a non-orthogonal lattice, and if $\omega$ 
is polygon in $\mathbb{L}$, then the subset of elementary 
moves in the BCC lattice which only involve the edges in 
the basis of $\mathbb{L}$ reduces to the usual simple 
cubic lattice moves.

By lemma \ref{lemma3} one can show that BCC polygons can be 
made contact free and so we only have to examine contact free
polygons in this section.

If $\omega$ is a contact free polygon in the BCC, then it is 
pushed into $\mathbb{L}$ by removing from it all edges in the
$\c2$ or $\c6$ directions.  The method of proof is as follows:
Replace very edge in the $\c2$ direction by the three edges 
$\c1\c4\c7$ and every edge in the $\c6$ direction by three 
edges $\c5\c8\c3$. 

All that remains is to show that one can arrange matters such
that self-intersections will not occur.

%%%%%%%%%%%%%%
\begin{theorem}
By applying the classes of BCC elementary moves in 
Figures~\ref{figa2} and~\ref{figa3} to unrooted polygons 
in the BCC lattice, any such polygon can be swept into a 
polygon in the sublattice $\mathbb{L}$.
\label{thm L} %%ZXZ[thm L]
\end{theorem}
%%%%%%%%%%%%%%

\Proof
Let $\omega$ be a contact free polygon and subdivide it 
twice in each of the $\c1$, $\c5$, $\c3$ and $\c7$ directions. 
Then the shortest distance in the $\c1$ direction between
vertices $\b{u}, \b{v}$ not connected by an edge of the 
polygon is at least $3$ steps.
Similarly for the $\c5$, $\c3$ and $\c7$ directions.

We now show that edges in line segments in the $\c2$ or 
$\c6$ directions can be swept into the sublattice $\mathbb{L}$. 

Replace edges in the $\c2$ or $\c6$ directions as follows:
\begin{itemize}
 \item If $\c2\c2\c2\ldots\c2$ is a sequence of consecutive 
edges in the $\c2$ direction, then replace them by 
$\c1\c4\c7\c1\c4\c7\ldots \c1\c4\c7$.
 \item Similarly, replace any sequence of consecutive edges 
$\c6\c6\c6\ldots\c6$ by the sequence 
$\c5\c8\c3\c5\c8\c3\ldots \c5\c8\c3$.
\end{itemize}

These changes can be achieved by positive non-planar BCC 
elementary moves on the BCC polygon. Observe that these
substitutions insert new vertices in the $\c1$, $\c5$ and
$\c7$ and $\c3$  directions, adjacent to existing vertices 
in $\omega$, but that no contacts can be created since 
other vertices in these directions are a distance of at
least three away.

The only intersections that can arise in the above 
construction do so at the endpoints of a sequence 
$\c2\c2\c2\ldots\c2$ or $\c6\c6\c6\ldots\c6$. These are avoided
as follows:
\begin{itemize}
 \item 
If the edges at the beginning of the sequence 
$\c2\c2\c2\ldots\c2$ is $\c5\c2\ldots$, then the substitution 
$\c2\to\c1\c4\c7$ on the first $\c2$ will cause an
intersection (or a ``spike'') $\c5\c1\c4\c7\ldots$. If 
instead a neutral non-planar elementary move 
$\c5\c2 \to \c4\c7$ is executed here, then the spike is 
avoided while the edge in the $\c2$ direction is removed.
\item 
A similar argument holds if the sequence of $\c2$'s 
ends as $\ldots \c2\c3$, in which case one finds $\c2\c3 
\to \c1\c4$.
\item 
Similar arguments can be used to deal with the case of 
line segments of the form $\c6\c6\ldots\c6$.
\item 
Lastly, if the line-segment has length one and is of the form 
$\c5\c2\c3$, then one obtains the negative BCC elementary 
move $\c5\c2\c3 \to \c5\c1\c4\c7\c3 \to \c4$.
\item 
A similar argument holds for segments of the form $\c1\c6\c7$.
\end{itemize}

Completion of these elementary moves produces a polygon with 
no edges in the $\c2$ or $\c6$ direction. This is exactly 
a polygon in the sublattice $\mathbb{L}$. This completes
the proof. \qed

\subsection{Irreducibility classes of the BCC elementary moves}

By Theorem~\ref{thm L} all BCC lattice polygons can be 
changed into lattice polygons in the lattice $\mathbb{L}$ 
which is isotopic to the simple cubic lattice. Since this
process is reversible, one only has to consider polygons
in $\mathbb{L}$ and the effects of the BCC elementary moves 
on these polygons in this sublattice.

Restricting the BCC elementary moves to the sublattice 
$\mathbb{L}$ implies that all elementary moves including 
either the directions $\c2$ or $\c6$ must be excluded.  
Observe that if both $\c2$ or $\c6$ are excluded from the 
set of BCC elementary moves (for example, 
$\c1\leftrightarrow\c2\c1\c6$ or $\c1\leftrightarrow \c2\c3\c8$, 
are excluded) then the remaining moves reduce to the 
standard simple cubic lattice BFACF moves in Figure~\ref{figa4}
in $\mathbb{L}$. 

Thus if we restrict ourselves to this subset of possible moves 
on polygons in $\mathbb{L}$ then we have the following 
lemma, which is a direct corollary of Theorem~\ref{thm 11} 
or Theorem~3.11 in reference~\cite{JvRW91}.

\begin{lemma}
The BCC elementary moves, restricted to the sublattice 
$\mathbb{L}$, applied to unrooted polygons in $\mathbb{L}$, 
have irreducibility classes which coincides with the
knot types of the polygons as piecewise linear embeddings 
in $\Rm^3$. \qed
\end{lemma}

Since every unrooted polygon in the BCC lattice can be (reversibly) reduced to a polygon in $\mathbb{L}$, the 
following theorem is an immediate corollary of 
Theorem~\ref{thm L} and the last lemma.

\begin{theorem}
\label{thm BCC main}
The irreducibility classes of the BCC elementary moves, applied
to unrooted polygons in the BCC lattice, coincides with the knot
types of the polygons as piecewise linear embeddings in 
$\Rm^3$. \qed
\end{theorem}

This completes the proof.  In other words, it follows that
the set of BCC elementary moves, applied to unrooted 
polygons in the BCC lattice, is irreducible within the knot 
type of the polygon.

%%%%%%%%%%%%%%%%%%%%%%%%%%%%%%%%%%%%%%%%%%%%%%%%%%%%%%%%%%
%%%%%%%%%%%%%%%%%%%%%%%%%%%%%%%%%%%%%%%%%%%%%%%%%%%%%%%%%%
%%%%%%%%%%%%%%%%%%%%%%%%%%%%%%%%%%%%%%%%%%%%%%%%%%%%%%%%%%
\section{BFACF-Style Elementary Moves in the FCC Lattice}
\label{section4}

In this section we introduce local elementary moves of a 
BFACF-style algorithm on the FCC lattice; in fact there is 
only a single move that replaces two edges by a single edge and
vice-versa. As was done for the BCC moves described above, 
we show that this new move is sufficient to realise a 
piecewise linear orientation preserving isotopy between two
unrooted FCC polygons of the same knot type. This demonstrates 
that the irreducibility classes of the FCC elementary 
move coincide with the knot types of unrooted FCC polygons
as determined by their embeddings in three space.

Local elementary moves in the FCC lattice have been explored 
previously in the literature, see for example references 
\cite{DCK86}, however, the ergodicity properties of these
elementary moves in the FCC have not been studied for knotted
polygons.

\begin{figure}
 \begin{center}
  \includegraphics[height=6cm]{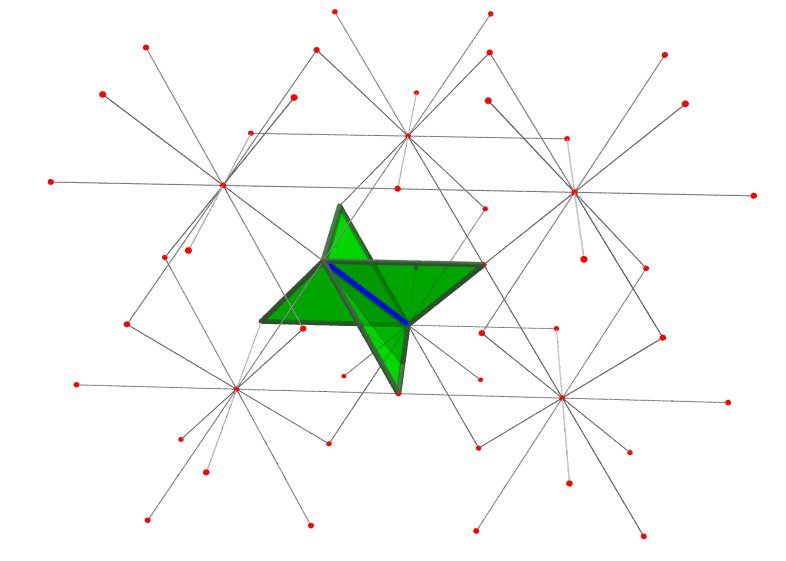}
 \end{center}
  \caption{The four plaquettes adjacent to an edge in the FCC lattice.}
  \label{fig fcc plaquettes}
\end{figure}

\subsection{The FCC elementary move}
\label{section41}

Vertices of the face centered cubic (FCC) lattice are points 
in $\Rm^3$ with positions given by the linear combinations 
$p\d1 + q\d2 + r\d3 + s\d4 + t\d5 + u\d6$ where
$p,q,r,s,t,u \in \Zm$, and where the vectors $\d{j}$ is given
\begin{displaymath}
\begin{array}{lll}
\d1 = (1,1,0),&\d2 = (1,-1,0),& \d3=(1,0,1),\\
\d4 = (1,0,-1),& \d5 = (0,1,1),& \d6=(0,1,-1),\\
\d7 =(-1,-1,0),& \d8 = (-1,1,0),& \d9 = (-1,0,-1),\\
\dd10 = (-1,0,1),& \dd11=(0,-1,-1),& \dd12 =(0,-1,1)
\end{array}
\end{displaymath}
We have labelled these vectors so that $\mathbf{d}_{i+6} 
= - \mathbf{d}_i$. We define \textit{adjacent}, 
\textit{lattice edges}, \textit{end-vertices}, 
\textit{lattice polygon} and \textit{line segment} on 
the FCC lattice in the same way that we did on the
BCC lattice. Note that a polygon of $n$ edges on the FCC 
has geometric length $n \sqrt{2}$.

Observe that the sublattice generated by any set of 
three different non-coplanar vectors in the generating set of 
the FCC is isotopic to the simple cubic lattice. For example, 
the set $\{\d1,\d3,\d5\}$ generates a sublattice of the FCC
which is isotopic to the cubic lattice (and may be viewed as 
a non-orthogonal simple cubic lattice).

Similarly, there are sets of three coplanar vectors which 
generate a two-dimensional triangular lattice; for example,
the set $\{\d1,\d3,\dd12\}$ are the basis vectors of a 
triangular lattice in a plane with normal vector 
$\d1\times \d3$ (and observe that $\dd12=\d3-\d1$).

\noindent{\U{FCC Elementary Move:}}
\begin{itemize}
\item
See Figure~\ref{fig fcc plaquettes}. Any vector
$\d{i}$ in the FCC lattice is incident to four triangular 
plaquettes. That is, there are four different pairs of 
vectors $(\d{j},\d{k})$ so that $\d{i} = \d{j}+\d{k}$. The
substitution of $\d{i}$ by $\d{j}\d{k}$ is a {\it positive 
atmospheric FCC move}, or a {\it positive FCC elementary move}
and it increases the length of the polygon by one edge. 
Similarly the reverse move, replacing $\d{j}\d{k}$ by $\d{i}$ 
is a {\it negative atmospheric FCC move} or a {\it negative 
FCC elementary move} and it decreases the length of the 
polygon by one.
\end{itemize}

\subsection{Stretching polygons by using the FCC elementary move}

In examining the irreducibility classes of the FCC elementary 
move, we shall follow the same strategy used for the BCC 
lattice in Section~3.  

We first show that FCC polygons can be stretched, then 
made contact free, and finally swept into a simple cubic 
sublattice of the FCC. In this simple cubic sublattice of 
the FCC, the usual simple cubic lattice BFACF moves can 
be performed on the polygon by using combinations of the
FCC elementary moves. Theorem~\ref{thm 11} can then be used 
to complete the proof that the irreducibility classes of 
the elementary move coincides with the knot types of
the polygon.

%%%%%%%%%%%%%%%%%
\begin{figure}[t]
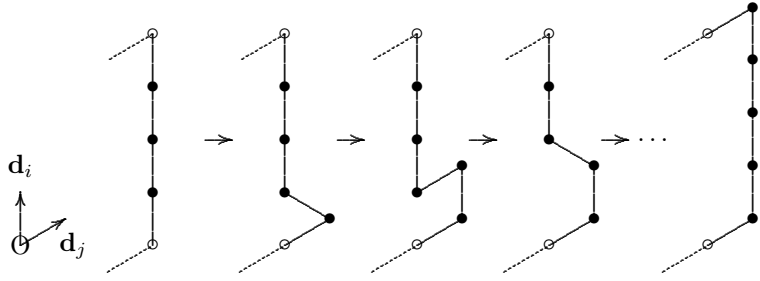

\input figu1.TEX
\caption{Translating a line segment in the $\d{i}$ direction
in a polygon in the $\d{j}$ direction using the elementary
move in the FCC lattice.}
\label{figu1} %%ZXZ[figu1]
\end{figure}
%%%%%%%%%%%%%%%%

The basic construction in the proof is illustrated in 
Figure~\ref{figu1}. By using the elementary move, the 
construction in Figure~\ref{figu1} shows that one can translate 
an entire line segment (in the $\d{i}$ direction one step in 
the $\d{j}$ direction, provided that $\d{j} \not= \pm \d{i}$.

This construction can be performed regardless of the 
orientation of the edges incident at the ends of the line 
segments, provided that the set of target vertices in the 
$\d{j}$ direction are not occupied by the polygon.

Let $\{ \d{i}, \d{j} , \d{k} \}$ be a triple which form a 
triangular plaquette in the FCC lattice (so $\d{k}=\d{j}-\d{i}$).
Note that these three vectors generate a triangular sublattice 
$T$ in the FCC; this sublattice is a geometric plane $A$ with
normal $\d{i}\times \d{j}$.

Consider the (non-orthogonal) projection of an FCC polygon 
$\omega$ into the plane $A$ along a lattice direction $\d{l}$ 
which is transverse to the plane $A$ (that is, $\d{l}$ is 
not coplanar with vectors $\{ \d{i},\d{j},\d{k}\}$). We say 
that this projection is taken along $\d{l}$ into the 
$\d{i}\d{j}$-plane $A$.

\begin{figure}[t]
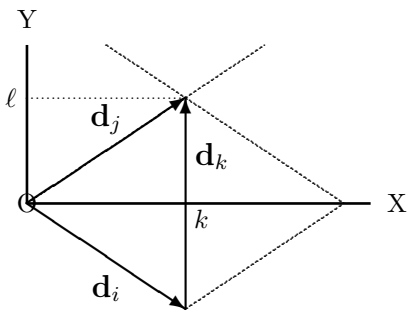

\input FIG1.TEX
\caption{
The vectors $\{ \d{i}, \d{j}, \d{k} \}$ generate a 
triangular lattice in the plane $A$. The Cartesian coordinate
system $(X,Y)$ is set up with $X$-direction given by
$\d{i}+\d{j}$ and $Y$-direction given by $\d{k}$.  Observe that 
$|\d{n}|=\sqrt{2}$, so that $k = \sqrt{3/2}$ and 
$\ell=1/\sqrt{2}$.}
\label{FIG1} %%ZXZ[FIG1]
\end{figure}

In general, the projection of an FCC lattice polygon $\omega$ 
along $\d{l}$ into the plane $A$ is a lattice knot projection 
into a triangular lattice which is the projection of the FCC 
onto $A$. Define a two dimensional cartesian coordinate system 
in the plane A with $Y$-direction given by $\d{k}$ and 
$X$-direction by $\d{i}+\d{j}$.  This is illustrated in 
Figure~\ref{FIG1}.

%%%%%%%%%%%%%%%%%
% %%%%%%%%%%%%%%%%

If $P\omega$ is the projection of the polygon $\omega$ in 
the plane $A$ along a transverse lattice direction $\d{l}$, 
then $P\omega$ is a finite graph in the triangular lattice
in $A$.  

Hence, there exists a right-most line $K$ parallel to the 
$Y$-axis (see Figure~\ref{FIG1}) which intersects the projection 
$P\omega$.  We say that the line $K$ is the \textit{maximal} 
or \textit{right-most} line {\it cutting} the projection 
$P\omega$. The right-most line is itself the projection of a plane $P_r$ projected with $\omega$ along the (non-orthogocal)
directopm $\d{l}$. The intersection of this plane with the 
FCC lattice is a triangular sublattice of the FCC,
generated by the set of basis vectors $\{\d{k},\d{l}\}$.
  
Line segments in $\omega$ are projected to line segments 
or to points in $K$, and these projected images lift back 
up to parts of the polygon in the intersection $P_r\cap \omega$
of the plane and the polygon.  Each line segment in this 
intersection can be translated one step in the $\d{j}$ 
(or $\d{i}$) direction (transverse to the plane $P_r$) by 
using the basic constructions in Figure~\ref{figu1}. If
all the line segments in the plane which projects to the
maximal or right-most line $K$ are moved in the $\d{j}$
direction, then we say that the polygon is {\it stretched} in 
the $\d{j}$ direction from the plane $P_r$.

\begin{figure}[t]
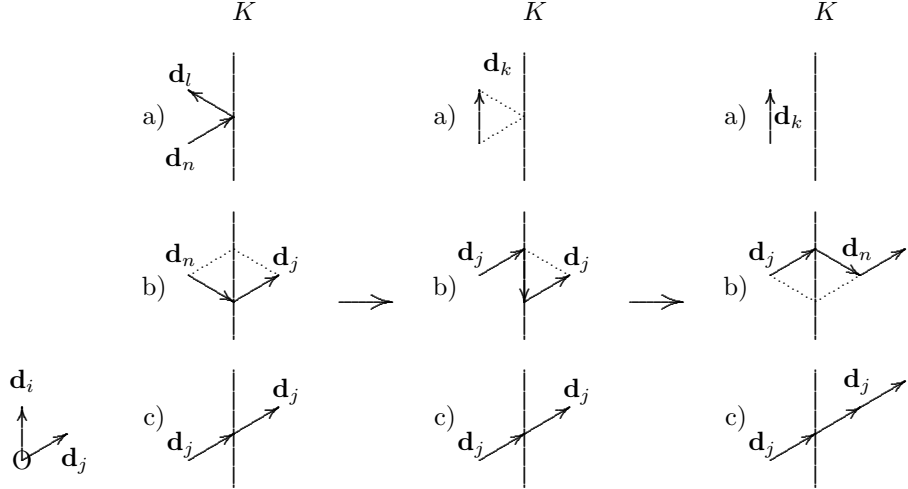

\input figu3.TEX
\caption{The projection of the polygon close to the 
maximal or right-most line $K$ during the stretching of
the polygon in the $\d{j}$ direction. The conformations of 
edges incident with the plane $P_r$ are in general one of 
the cases above.  In two cases the polygon passes through the
$P_r$.  In the third case it touches it in one vertex 
before turning back. In this last case the vectors $\d{n}$ 
and $\d{l}$ must make an angle of $60^o$ with one another.
Cases as in (a) are removed as illustrated, while case (c)
is left unchanged.  In case (b) the edge in the $\d{n}$
direction is passed through the plane $P_r$ in the direction
$\d{j}$. See Figure~\ref{figu4} for more details.}
\label{figu3} %%ZXZ[figu3]
\end{figure}

Observe that once a polygon has been stretched in the plane 
$P_r$, then there are no line segments in the polygon 
contained in $P_r$.  This, in particular, implies that 
$P_r\cap \omega$ is a collection of isolated vertices
in $\omega$.  At each such vertex $\omega$ either passes 
through the plane to its right in the $\d{j}$ direction, or 
it turns to stay to the left of $P_r$.  Projected images 
are illustrated in Figure~\ref{figu3}.

%%%%%%%%%%%%%%%%%
%%%%%%%%%%%%%%%%

Next we stretch the polygon in a plane $P_r$ recursively 
starting in its right-most line $K$, and then successively 
moving left.  Line segments in $P_r \cap \omega$ can be
moved in the $\d{j}$ direction using the construction
in Figure~\ref{figu1}.  Since all the edges incident with
$P_r$ on the maximal side of $P_r$ are in the $\d{j}$
direction, these constructions can be performed without
creating any self-intersection.

This leaves the cases of isolated vertices in $P_r\cap\omega$.
The situation is as illustrated in Figure~\ref{figu3}:  All 
the parts of  the polygon to the right of $P_r$ $K$
been stretched in the $\d{j}$ direction, and the next step 
is to move parts of the polygon which has isolated vertices
in $P_r$ in the $\d{j}$ direction.

In Figure~\ref{figu3}(a) we depict a situation in which 
the two vectors $\a{n}$ and $\a{l}$ must make a $60^o$ angle 
with one another at the vertex in $P_r$.  In this case
we can remove these two edges from the polygon by making a 
negative atmospheric move replacing them with a single edge 
as shown. Thus we can eliminate this situation and we do
not need to consider it in the discussion below.

\begin{figure}[t]
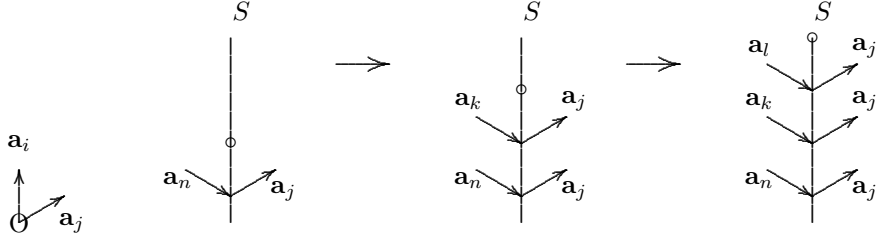

\input figu4.TEX
\caption{The case in Figure~\ref{figu3}(b). If $Q$ is a plane
coplanar with $\{\d{j},\d{n}\}$ then $S$ is the line $S 
= Q\cap P_r$.  If one moves along $S$ in the vertical sense
here, then eventually one must encounter an open vertex
$\circ$.  In this case the moves in Figure!\ref{figu3}(b)
can be systematically performed on each of the isolated
vertices along $S$, starting at the top.  This will translate
all the edges on the left of $S$ through the plane $P_r$ 
towards the right in the $\d{j}$ direction.}
\label{figu4} %%ZXZ[figu4]
\end{figure}

Next, we address the situation depicted in 
Figure~\ref{figu3}(b). We need to translate the edge $\d{n}$ 
in the $\d{j}$ direction.  This is done as illustrated in
Figure~\ref{figu4} (see also Figure~\ref{figu3}). If the 
vertex marked by $\circ$ in the left-most figure in 
Figure~\ref{figu4} is vacant (that is, not occupied by 
the polygon), then we can use two elementary moves to change 
the conformation $\d{n}\d{j}$ to $\d{j}\d{n}$; so the vector 
$\d{n}$ is translated in the $\d{j}$ direction, as desired.

On the other hand, if the vertex marked by $\circ$ is occupied 
by another part of the polygon, then there are only two
possibilities. First, it may correspond to the conformation
depicted in Figure~\ref{figu3}(a), in which case the 
elementary move can be used to remove the occupied vertex. 

Otherwise it is the situation shown in the middle of
Figure~\ref{figu4}. In this case let $Q$ be the plane
coplanar with $\{\d{j},\d{n}\}$ and consider the line
$S = Q\cap P_r$.  Move along $S$ in $P_r$ until an open vertex 
$\circ$ is enountered as shown the second and third parts 
of Figure~\ref{figu4}.

Since the polygon is finite we eventually must find a 
vacant vertex --- as shown in the rightmost past of 
Figure~\ref{figu4}. The edge (depicted as $\d{l}$) can 
then be translated in the $\d{j}$ direction. This 
(effectively) moves the vacant vertex back along $P_r$ and 
so one recursively apply this construction until we finally 
move the desired edge, $\d{n}$, in the $\d{j}$ direction,
using the construction in Figure~\ref{figu3}(b).

%%%%%%%%%%%%%%%%%
%%%%%%%%%%%%%%%%

The stretching of the polygon in a new plane, $Q_r$, 
in the transverse (to $Q_r)$ direction $\d{j}$, proceeds 
by first finding the right most line in the projection,
and lifting it to the plane $P_r$ (parallel to $Q_r$). The 
polygon is then stretched one step in the $\d{j}$ direction 
in $P_r$.  Then $P_r$ is moved closer to $Q_r$ one 
step in the $-\d{j}$ direction and the polygon is stretched
recursively until it is finally $P_r$ is coincident with 
$Q_r$ in which case the polygon is stretched one step in 
the $\d{j}$ direction on one side of $Q_r$.

The effect of this construction is to cut the polygon 
along the plane $Q_r$ into a left part and right part. 
The right part is then translated in the $\d{j}$ direction 
and the two parts are reconnected by inserting inserting edges 
in the $\d{j}$ direction.

In topological terms this construction is an (orientation
preserving) ambient isotopy of three space, and it does 
not change the knot type of the polygon.  We say that the 
polygon $\omega$ was {\it stretched} in the $\d{j}$ 
direction transverse to the plane $Q_r$.

This construction proves the following lemma:

%%%%%%%%%%%%%
\begin{lemma}
Let $\omega$ be any FCC lattice polygon and let $Q_r$ 
be a lattice plane intersecting $\omega$.  Then $\omega$ 
can be stretched, using the FCC elementary move, in $Q_r$ in
the $\d{j}$ direction if $\d{j}$ is transverse to $Q_r$ .
\label{lemma 7} %%ZXZ[lemma 7]
\end{lemma}
%%%%%%%%%%%

\Proof
Since $Q_r$ is a lattice plane, determine a plane $A$ 
transverse to it which intersects the FCC in a triangular
lattice.  Orient he projected polygon as above, and then
use the elementary moves to stretch the polygon in $Q_r$ in 
the desired direction as described by the construction 
above. \qed

%%%%%%%%%%%%%%%%%%%%%%%%%%%%%%%%%%%%%%%%%%%%%%
\subsection{Contact free FCC lattice polygons}

As was the case for the BCC lattice, care must be taken to 
ensure that different pieces of the polygon do not come 
too close together while we try to sweep it onto a simple
cubic sublattice. Thus we define \textit{contacts} on the 
FCC lattice in the same way we did on the BCC lattice; 
two non-adjacent vertices $\b{u}$ and $\b{v}$ in a polygon
$\omega$ form a {\it contact} if $\b{u}-\b{v} = \d{j}$ for 
some $j$.  That is, these vertices are a distance 
$\sqrt{2}$ apart in the FCC.  

A polygon $\omega$ is {\it contact free} if it has no 
contacts. We next show that we can use the FCC elementary 
move to transform any polygon into a contact free polygon.

Lemma \ref{lemma 7} shows that any given polyon can be 
stretched in the $\d{j}$ direction transverse to a plane 
$Q_r$ (which intersects the FCC lattice in a 
triangular sublattice). 

If a polygon is stretched in every transverse direction
in every lattice plane $Q$ which intersects it in a 
triangular lattice, then the effect is to double the length
of every edge of the polygon.  That is, each edge gets
replaced by two edges.  This is a {\it subdivision} of 
the polygon $\omega$. One may repeat this stretching several
times to stretch each edge any number of times.

If polygon has a contact in the $\d{j}$ direction,
then it is removed if the polygon is subdivided once since
all distances between vertices in the polygon is increased
by at least a factor of $2$ in a subdivision of
the polygon.  In addition, a contact in the
$\d{j}$ direction cannot be created if the polygon is
stretched in the $\d{i}$ direction, since the relative
orientations of vertices and edges in the $\d{j}$ direction
is maintained if the stretching is in the $\d{i}$
direction.  

Thus, one may remove all contacts from a
polygon by stretching or by subdivision.  This leaves
a contact free polygon, and we proved the following lemma.

%%%%%%%%%%%%%
\begin{lemma}
By subdividing an FCC polygon, it is possible to transform 
it into a contact free polygon, using the FCC elementary move.
\label{lemma contact}
\end{lemma}
%%%%%%%%%%%%%
\Proof
Consider a (non-orthogonal) projection $P\omega$ of an 
FCC polygon $\omega$ into a lattice plane $A$ along
a lattice direction transverse to $A$, where $A$ intersects
the FCC in a triangular lattice $\Lm_t$ with basis vectors (say) 
$\d{i}$, $\d{j}$ and $\d{k}$. 

The projection $P\omega$ cuts the plane $A$ into one infinite
face and a set of finite faces (or areas). Each finite face 
has an area measured in units of the elementary triangle in 
the triangular lattice. 

Since $\d{i}$ is a basis vector of the sublattice $\Lm_t$,
a subdivision of $\omega$ in the $\d{i}$ direction increases 
the area of each finite face in the projection in the plane 
$A$ by at least one unit elementary triangle. This is 
similarly true when subdividing in the $\d{j}$ and 
$\d{k}$ directions.

Contacts in the FCC polygon which are oriented in the 
$\d{i}$, $\d{j}$ and $\d{k}$ directions will project in 
the plane $A$ as edges in $\Lm_t$.  A subdivision in the 
direction of a given contact will remove it from the 
projection, and thus also from the polygon itself.

If subdivisions in the $\d{i}$ direction removed all 
contacts in this direction, then subsequent subdivision in
other directions will not create contacts in the $\d{i}$
direction (since this would require the translation of 
parts of the polygon in the $\d{i}$ direction, and this 
cannot occur).  

Hence, by using subdivisions of $\omega$ in each of the
six independent directions of the FCC lattice, all contacts 
are removed from the polygon, and it becomes contact free. 
This completes the proof. \qed

\subsection{Pushing contact free FCC polygons into a simple cubic sublattice}

Let $\mathbb L$ be that sublattice of the FCC with basis
vectors $\{\d{1},\d{3},\d{5},\d{7},\d{9},\dd{11} \}$.
Then $\mathbb L$ is ambient isotopic to the simple cubic
lattice.

If $\omega$ is a polygon in $\mathbb L$, then the 
(simple cubic lattice) BFACF move (see Figure~\ref{figa4}) 
can be performed on $\omega$ in $\mathbb{L}$ by composing 
two FCC elementary moves in order to execute each single 
BFACF move.  For example, the positive BFACF move 
$\d{1} \to \d{3}\d{1}\d{9}$ can be performed in the two 
step sequence $\d{1} \to \d{3}\d6 \to \d{3}\d{1}\d{9}$.

In other words, the irreducibility properties of FCC
lattice polygons in the sublattice $\mathbb L$ is
determined by the irreducibility properties of the 
simple cubic lattice BFACF moves, and by Theorem \ref{thm 11}.
We shall use these facts to complete the proof that 
the irreducibility classes of the FCC elementary move 
on unrooted FCC lattice polygons coincide with the knot types 
of the polygons as piecewise linear embeddings in three space.

The strategy is as follows:  If $\omega$ is an FCC lattice
polygon, then it will be moved to a polygon in the
sublattice $\mathbb L$, from where it will be put in
standard position $\lambda$ given its knot-type $K$.
Since the composition of all the FCC elementary moves
is a continuous piecewise linear orientation preserving
isotopy of three space, we know that the polygon 
$\lambda$ in standard position has the same knot type
$K$, and moreover, that every polygon can be moved to
$\lambda$ using a finite number of FCC elementary moves.

It only remains to prove that every FCC lattice polygon can
be moved to a polygon in $\mathbb L$.  

\begin{lemma}
By using the FCC elementary move, any FCC lattice polygon 
$\omega$ can be swept into a sublattice $\mathbb{L}$ of the FCC, 
where $\mathbb{L}$ is ambient isotopic to the simple cubic
lattice.
\label{lemma C}
\end{lemma}

\Proof
Since every polygon can be made contact free, it is enough 
to assume that $\omega$ is contact free.  We next prove 
that every such contact free polygon can be moved into 
the lattice $\mathbb L$.

If $\omega$ is embedded in $\mathbb{L}$, then we are done.

Otherwise,  $\omega$ has some edges in directions from the 
set $\{\d{2}, \d{4},\d{6}, \dd{8},\dd{10},\dd{12} \}$. 
Since $\omega$ is contact free, the FCC elementary move 
can be performed on each of each of the edges in $\omega$
without creating self-intersections. 

Proceed along the polygon and remove edges in the 
${\bf d}_{10}$
$\d{2}$, $\d{4}$, $\d{6}$, $\d{8}$, ${\bf d}_{10}$ and 
${\bf d}_{12}$ directions systematically by replacing them 
as follows:
\begin{displaymath}
  \begin{array}{lll}
  \d2 \to \d3\dd{11}, &
  \d4 \to \d1\dd{11}, &
  \d6 \to \d9\d5, \\
  \d8 \to \d9\d{5}, &
  \dd10 \to \d7\d5, &
  \dd12 \to \d3\d7.
  \end{array}
\end{displaymath}
If during this process a contact is created, then the polygon 
is again swept in the $\{\d1,\d3,\d5\}$ directions.  
Each contact will have a non-zero projection along at least
one of these directions, and will then be removed 
without inserting any new edges in any directions apart 
from those in the simple cubic sublattice $\mathbb{L}$. 
Finally, these elementary moves will move $\omega$ 
into the sublattive $\mathbb L$. \qed

A direct corollary of this lemma is

\begin{theorem}
\label{thm FCC main}
The irreducibility classes of the FCC elementary move, 
applied to unrooted polygons in the FCC, coincides with the 
knot types of the polygons as piecewise linear embeddings 
in $\Rm^3$. 
\end{theorem}

\Proof
Observe that since $\mathbb{L}$ is a sublattice of the 
FCC which is isotopic to the simple cubic lattice, 
the irreducibility classes of the FCC elementary move 
applied to unrooted polygons in $C$ coincides with the 
knot types of the polygons as piecewise linear
embeddinds in $\Rm^3$. This follows directly from the 
observation above that all BFACF moves on a polygon 
in $\mathbb{L}$ can be induced by using the FCC 
elementary move, and from Theorem~\ref{thm 11} or 
Theorem~3.11 in reference \cite{JvRW91}.

By Lemma~\ref{lemma C} any unrooted polygon in the FCC 
can be (reversibly) swept into the sublattice $\mathbb{L}$ 
using the FCC elementary move. 

Standard BFACF elementary moves in the sublattic $\Lm$
can be performed by combining FCC elementary moves.
For exampl, the positive move $\d{1} \to \d{3}\d{1}\d{9}$
in $\Lm$ is a BFACF move, and it is performed in two
steps in the FCC by $\d{1} \to \d{3}{\bf d}_{12}
\to \d{3}\d{1}\d{9}$, since $\d{1} = \d{3}+{\bf d}_{12}$
and ${\bf d}_{12} = \d{1} + \d{9}$.  The other cases,
and neutral and negative SC BFACF moves can similarly 
be checked.

Hence, since the standard SC BFACF moves in Figure~\ref{figa4} 
can be performed on the polygon in $\mathbb{L}$, using 
composite FCC elementary moves, the theorem follows 
immediately. \qed

%%%%%%%%%%%%%%%%%%%%%%%%%%%%%%%%%%%%%%%%%%%%%%%%%%%%%%%%%%%%%
%%%%%%%%%%%%%%%%%%%%%%%%%%%%%%%%%%%%%%%%%%%%%%%%%%%%%%%%%%%%%
%%%%%%%%%%%%%%%%%%%%%%%%%%%%%%%%%%%%%%%%%%%%%%%%%%%%%%%%%%%%%
\section{Conclusions and implementation}
\label{section5}

In this paper we have shown that the BFACF elementary 
moves on simple cubic lattice polygons can be generalised 
to similar moves on the BCC and FCC lattices. For the BCC
lattice, we propose a set of four elementary moves (two 
are neutral, and two are positive/negative), while on the 
FCC lattice we propose a single (reversible) elementary
move (of type positive/negative).

We proved that, on the BCC and FCC lattices, these moves 
are suitable to give irreducibility classes of unrooted 
polygons that coincide with the knot types of the
polygons. This is, we have generalised Theorem~3.11 in 
reference \cite{JvRW91} to both the FCC and BCC lattices. 
For the BCC lattices, our method of proof relied on all four
proposed elementary moves and so we know that these moves 
are sufficient.  It is not clear that any of these 
are necessary, or that a smaller set of similarly 
defined elementary moves may be sufficient.

Using the proposed moves we have implemented the GAS 
algorithm \cite{JvRR09} to sample knotted polygons on both 
the FCC and BCC lattices. We give some results in
Table~\ref{table} and further numerical results will be published elsewhere.

\subsection{The GAS implementation of the BCC and FCC
elementary moves}

In the GAS implementation \cite{JvRR09} of the algorithm, 
lattice polygons of (fixed) knot type $K$ are sampled along a 
Markov Chain by executing the elementary (or atmospheric) 
moves on the polygons.  Let $\phi = \langle \phi_n \rangle
= \langle \phi_0,\phi_1,\phi_2,\ldots\rangle$ be a realisation
of a sequence started in the initial state $\phi_0$, which
is a polygon of length $\ell(\phi_0)$ and of knot type $K$.

A state $\phi_{n+1}$ from $\phi_n$ is generated as follows:
Let $A_+(\phi_n)$ be the set of all positive atmospheric
or elementary moves on the polygon $\phi_n$, and similarly,
$A_0(\phi_n)$ and $A_-(\phi_n)$ the set of all neutral and
negative atmospheric moves on $\phi_n$. These are the
positive, neutral and negative \textit{atmospheres} of
$\phi_n$.

Define the sizes of the of the positive, neutral and negative
atmospheres of $\phi_n$ by $a_+(\phi_n)$, $a_0(\phi_n)$
and $a_-(\phi_n)$.  If $\ell(\phi_n) = N$, then each of these
statistics has a mean value denoted by $\langle a_+\rangle_N$,
$\langle a_0 \rangle_N$ and $\langle a_- \rangle_N$.

Next, define a set of parameters of the GAS algorithm by
\begin{equation}
\beta_N = \frac{\langle a_- \rangle_N}{\langle a_+ \rangle_N}.
\end{equation}
With these definitions, one may now determine $\phi_{n+1}$.

The state $\phi_{n+1}$ is obtained from $\phi_n$ by selecting
with probability $P_+$ a positive atmospheric move, with
probability $P_0$ a neutral atmospheric move, and with
probability $P_-$ a negative atmospheric move, and executing it
on $\phi_{n}$ to obtain $\phi_{n+1}$.  In each case, once
a determination is made to execute a move, the move is selected
uniformly from amongst the available moves in each of the
sets $A_+(\phi_n)$, $A_0(\phi)$ and $A_-(\phi_n)$.

The probabilities $P_+$, $P_0$ and $P_-$ are given by
\begin{eqnarray*}
P_+ &=& \frac{\beta_N a_+(\phi_n)}{
      a_- (\phi_n)+ a_0(\phi_n) + \beta_N a_+(\phi_n)}; \\
P_0 &=& \frac{a_0(\phi_n)}{
      a_- (\phi_n)+ a_0(\phi_n) + \beta_N a_+(\phi_n)}; \\
P_- &=& \frac{a_1(\phi_n)}{
      a_- (\phi_n)+ a_0(\phi_n) + \beta_N a_+(\phi_n)},
\end{eqnarray*}
where $N = \ell(\phi_n)$ is the length of the polygon $\phi_n$,
and $\beta_N$ is estimated by updating it recursively
in the simulation.  

This implementation give a sequence $\phi=\langle \phi \rangle_M$
of $M+1$ states sampled from polygons of knot type $K$. 
The analysis of the data sequence depends on a weight $W(\phi)$
of the sequence, which is defined as follows:  Define
$\sigma(n,n+1) = -1$ if $\phi_n \to \phi_{n+1}$ via a positive
atmospheric move, and $\sigma(n,n+1) = 1$ if $\phi_n 
\to \phi_{n+1}$ via a negative atmospheric move.  Then
the weight of the sequence $\phi$ is given by
\begin{equation}
W(\phi) = 
\left[ \frac{a_-(\phi_0)+a_0(\phi_0)
           +\beta_{\ell(\phi_0)} a_+(\phi_0)}{              
             a_-(\phi_L)+a_0(\phi_L)
           +\beta_{\ell(\phi_L)} a_+(\phi_L)} \right] 
\prod_{n=0}^{|\phi|-1} \beta_{\ell(\phi_n)}^{\sigma (n,n+1)} , 
\end{equation}
where $\ell(\phi_n)$ is the length (number of edges) in
polygon $\phi_n$, and where $L=|\phi|$ is the length (number of
states) in the sequence $\phi$.

The expected value of the weight of sequences ending in
states of length $N$ is $\langle W (\phi) \rangle_N$, and the
basic GAS theorem states that
\begin{equation}
\frac{\sum_{|\tau| = N} \langle W(\tau) \rangle_N}{
\sum_{|\rho| = M} \langle W(\rho) \rangle_M} 
= \frac{p_N(K)}{p_M(K)},
\end{equation}
where $\tau$ are all possible sequences of states ending
in a state of length $|\tau| = N$, and $\rho$ are
all the possible sequences of states ending in a state
of length $|\rho| = M$.

In this implementation, the FCC and BCC elementary moves
described in this paper will give an approximate
enumeration algorithm, and estimates of the number of
polygons of knot type $K$ and length $n$, $p_n(K)$, can
be obtained.  This approach produced the data in table
\ref{table}.

\subsection{The BFACF-style implementation of the BCC and
FCC elementary moves}

The GAS algorithm is by no means a standard simulation 
method (though the authors hope that it might become 
more popular), and so we finish the paper by discussing 
how the proposed moves may be implemented in a 
BFACF-style algorithm, as implemented in reference
\cite{ACF83,BF81}; further implementations can be found
in references \cite{BN97,OJvRW96}.

In general the BFACF algorithm is applied to either a self-avoiding walk or self-avoiding polygons and so we 
will discuss the implementation of the BCC and FCC moves 
to sample either walks or polygons. Since these moves do 
not translate the endpoints of a self-avoiding walk, assume 
that the walk is rooted at its endpoints. The simple cubic
lattice moves are not irreducible on the state space $S_{0x}$ 
of walks with fixed endpoints $0$ (the origin) and a lattice 
site $x$ \cite{JvR92,JvRW91} unless those endpoints are a 
distance of at least 2 apart (for details, see reference 
\cite{JvR92}). We believe that a similar result will hold 
for the BCC and FCC moves we have proposed, but we have not yet
investigated this.

Let $\omega_n$ be the current walk or polygon (or ``state'') 
of length $|\omega_n|$ edges. Choose an edge $s$ from 
$\omega_n$ with uniform probability. Enumerate the possible
elementary moves on $s$; in the BCC lattice there are $12$ 
possible moves while in the FCC there are $4$ possible moves.  
At most one these elementary moves is a negative atmospheric 
move (will shorten the polygon or walk), while the remaining 
moves are positive or neutral.

Choose one of the possible moves so that a particular positive 
move is done with probability $P_+$, a particular negative move 
is carried out with probability $P_-$ and a particular neutral 
elementary move is performed with probability $P_0$.  This
produces a state $\omega_n^\prime$. If this state is self-avoiding 
then we accept it and set $\omega_{n+1} = \omega_n^\prime$ 
and otherwise we reject it and set $\omega_{n+1} =
\omega_n$. It remains to specify the probabilities $P_+$, $P_-$ 
and $P_0$, and hence a probability is attached to each of 
these moves. This must be done differently on each lattice.

Consider a polygon or walk in the BCC lattice and a given edge 
$s$. The possible moves we can perform on $s$ depend on the 
conformation of the of $s$ and its predecessor and successor 
edges. In particular, the moves on a given edge are in one 
the following classes:
\begin{itemize}
\item All the possible moves on the edge $s$ are 
positive elementary moves increasing the length of the walk 
or polygon; this gives $12$ positive elementary moves on $s$.
\item The possible elementary moves include $11$ positive 
and $1$ neutral move; denote this as $(11,1,0)$.
\item The possible elementary moves include $10$ positive 
and $2$ neutral moves; denoted by $(10,2,0)$.
\item The possible elementary moves include $9$ positive, 
$2$ neutral and $1$ negative moves; denoted by $(9,2,1)$.
\item Similarly, the following combinations are possible 
$(9,3,0)$, $(8,4,0)$, $(7,4,1)$ and $(6,6,0)$.
\end{itemize}

In addition, one has the Boltzmann type relation $P_+ = 
e^{-2\beta} P_-$ relating the probability of making a 
positive elementary move, to that of making a negative
elementary move. The cases above sets up a set of bounds on 
the different probabilities:
% \begin{eqnarray}
% 12 P_+ &\leq 1; \nonumber \\
% 11 P_+ + P_0 & \leq 1; \nonumber \\
% 10 P_+ + 2P_0 & \leq 1; \nonumber \\
% 9 P_+ + 3 P_0 & \leq 1; \nonumber \\
% 9 P_+ + 2 P_0 + P_- & \leq 1; \nonumber \\
% 8 P_+ + 4 P_0 & \leq 1; \nonumber \\
% 7 P_+ + 4 P_0 + P_- & \leq 1; \nonumber \\
% 6 P_+ + 6 P_0 & \leq 1. \nonumber
% \end{eqnarray}
\begin{displaymath}
 \begin{array}{llcll}
  12 P_+ &\leq 1; &\qquad & 11 P_+ + P_0 & \leq 1; \\
  10 P_+ + 2P_0 &\leq 1; && 9 P_+ + 3 P_0 & \leq 1; \\
  9 P_+ + 2 P_0 + P_- & \leq 1; && 8 P_+ + 4 P_0 & \leq 1; \\
  7 P_+ + 4 P_0 + P_- & \leq 1; && 6 P_+ + 6 P_0 & \leq 1.
 \end{array}
\end{displaymath}
We need to maximise the value of $P_+$ while satisfying 
these bounds . Since $\beta>0$ and $P_+ = e^{-2\beta} P_-$,  
it follows that $P_+ \leq P_-$. If we additionally assume 
that $P_+ \leq P_0 \leq P_-$, then the first six inequalities 
are redundant and one can simply study the last two 
(that is, the bottom row). Maximising then gives
\begin{equation}
\hspace{-1cm}
P_+ = \frac{e^{-2\beta}}{3(1+3e^{-2\beta})},\quad
P_0 = \frac{1+e^{-2\beta}}{6(1+3e^{-2\beta})},\quad
P_- = \frac{1}{3(1+3e^{-2\beta})}.
\end{equation}
Note that $P_0 = \frac{1}{2}(P_++P_-)$.

The situation is simpler on the FCC lattice due to the 
simplicity of the elementary move. When an edge $s$ is 
selected from the polygon there are only two possibilities:
\begin{itemize}
\item the four possible moves on the edge $s$ are 
positive elementary moves
\item three of the possible elementary moves are positive and 
one is a negative elementary move.
\end{itemize}
This gives the following set of inequalities for the probabilities $P_+$, and $P_-$:
\begin{displaymath}
4P_+ \leq 1; \qquad\qquad 3P_+ + P_1 \leq 1.
\end{displaymath}
Since there are no neutral moves in the FCC algorithm 
we do not require $P_0$ (or we can simply set $P_0=0$). 
Again we have a Boltzmann factor relating the probabilities, 
namely $P_+ = e^{-\beta} P_-$ with $\beta>0$. Since 
$P_+ \leq P_-$ the first of these equations is redundant, 
and the second gives
\begin{equation}
% \hspace{-1cm}
P_+ = \frac{e^{-\beta}}{1+3e^{-2\beta}},\quad\qquad
P_- = \frac{1}{1+3e^{-\beta}}.
\end{equation}

With the above choices of $\{P_+,P_0,P_-\}$ on either the 
BCC or FCC lattices, the algorithm for polygons runs 
as described below. The algorithm for walks is identical after
the initialisation step. 
\begin{itemize}
\item[\bf 1.] Initialise the algorithm by choosing the 
first polygon, $\omega_0$, to be a polygon of given length 
and desired knot type $K$. Define the parameter $\beta$. The
length of this initial polygon is not crucial except that it 
must be sufficiently long to embed a knot of type $K$.
\item[\bf 2.] Let $\omega_n$ be the current polygon of 
length $|\omega_n|$ edges.  With uniform probability, 
$1/|\omega_n|$, choose an edge $s$ in $\omega_n$.  Enumerate all
possible elementary moves which can be performed on that edge.
\item[\bf 3.] With probability $P_+$ perform a (particular) 
positive atmospheric move, uniformly selected from amongst 
all the possible positive moves.  Similarly, with
respectively probabilities $P_0$ or $P_-$, perform a 
particular neutral or negative move. If the sum of 
these probabilities is less than $1$, then the remaining 
default move is to leave the polygon unchanged.
\item[\bf 4.] Step {\bf 3} induces a proposed polygon 
$\omega_n^\prime$.  If the proposed polygon is not 
self-avoiding, or if (by default) no elementary move 
occurred, then $\omega_n^\prime = \omega_n$.
\item[\bf 5.] Put $\omega_{n+1} = \omega_n^\prime$ to find 
the next state, and continue at step {\bf 2} above.
\end{itemize}

In this implementation the transition probability of obtaining a polygon $\tau$ from a
given polygon $\omega$, $P_{\omega\to \tau}$, is given by
\begin{equation}
P_{\omega \to \tau} = \cases{
[P_+/|\omega|], & if the elementary move is positive; \cr
[P_-/|\omega|], & if the elementary move is negative; \cr
[P_0/|\omega|], & if the elementary move is neutral,}
\end{equation}
since the edge $S$ is selected with a priori probability
$1/|\omega|$ before the elementary move is performed. These
transition probabilities satisfy the condition of detailed
balance between states $\omega$ and $\tau$ given by
\begin{equation}
|\omega|\, P_{\omega\to \tau} = |\tau| P_{\tau\to \omega} ,
\label{eqndetbal}
\end{equation}
with the result that, in the BFACF-style implementation, 
the algorithm will sample asymptotically from the distribution

\begin{equation}
P_\beta (\omega) = \frac{|\omega|e^{\beta |\omega|}}{\sum_\tau |\tau|e^{\beta|\tau|}}
= \frac{|\omega|e^{\beta |\omega|}}{\sum_n n\,p_n (K) e^{\beta n}},
\label{eqnBFACFdist}  %%ZXZ[eqnBFACFdist]
\end{equation}
where $p_n (K)$ is the number of polygons of length $n$ 
edges and knot type $K$, since the algorithm is ergodic 
on the class of unrooted polygons with fixed knot type $K$.
This is not the canonical Boltzmann distribution over the 
state space of unrooted polygons with knot type $K$ given by
\begin{equation}
P_\beta^\prime (\omega) =  
\frac{e^{\beta |\omega|}}{\sum_\tau |\tau|e^{\beta|\tau|}}
= \frac{e^{\beta |\omega|}}{\sum_n n\,p_n (K) e^{\beta n}}.
\end{equation}
Note that denominator is finite
when $\beta < - \log \mu_K$ where $\mu_K$ is the growth 
constant of polygons of given knot type defined by 
\begin{equation}
\limsup_{n\to\infty}\left[ p_n (K) \right]^{1/n} = \mu_K.
\end{equation}
This limit is known to exist for the unknot, but its existence 
for other knot types is a long standing open problem.

\subsection{Metropolis-style implementation of BCC and FCC elementary moves}

A simpler implementation of the BCC and FCC elementary 
moves uses the Metropolis algorithm. In this case steps 
{\bf 2} and {\bf 3} above are modified as follows:  Choose
an edge $s$ with uniform probability $1/|\omega_n|$ in 
the current polygon, and select (uniformly) one of the 
$12$ possible elementary plaquettes on the edge in the 
BCC lattice, or one of $4$ possible elementary plaquettes 
on it in the FCC lattice. Then choose one of the 
elementary plaquettes uniformly and attempt the 
elementary move it defines.  If the resulting object 
is self-avoiding polygon $\omega_n^\prime$ of length
$|\omega_n^\prime|$ edges, then accept it with 
probability $\min\{1,e^{\beta (|\omega_n^\prime| 
- |\omega_n|)}\}$ as the next state $\omega_{n+1}$.  

Otherwise, put $\omega_{n+1} = \omega_n$.  In 
this implementation one can show that the condition 
of detailed balance is given by equation by 
equation~\Ref{eqndetbal}, and if $\beta < - \log \mu_K$, 
the algorithm samples asymptotically from the distribution 
$P_\beta (s)$ in equation~\Ref{eqnBFACFdist}.

\section*{Acknowledgments}
The authors acknowledge funding in the from of Discovery
Grants from NSERC (Canada).

\vspace{5mm}

\vspace*{0.5cm}

\noindent{\bf{Bibliography}}

\vspace*{0.3cm}

\begin{center}

\end{center}

% Set the ending of a LaTeX document
\end{document}